\documentclass[twocolumn,preprintnumbers,amsmath,amssymb]{revtex4}

\usepackage{graphicx}
\usepackage{dcolumn}
\usepackage{bm,epsfig}
\usepackage{epstopdf}
\usepackage{xcolor}

\begin{document}

\title{Single-component superconducting state in UTe$_{2}$ at 2~K}

\author{P. F. S. Rosa, A. Weiland, S. S. Fender, B. L. Scott, F. Ronning, J. D. Thompson, E. D. Bauer, and S. M. Thomas}
\affiliation{
Los Alamos National Laboratory, Los Alamos, New Mexico 87545, U.S.A.}
\date{\today}


\maketitle

\textbf{UTe$_{2}$ is a newly-discovered unconventional superconductor
wherein multicomponent topological superconductivity is anticipated based on the presence of two superconducting transitions and 
time-reversal symmetry breaking in the superconducting state~\cite{Ran2019,Hayes2020}.
The observation of two superconducting transitions, however, remains controversial~\cite{Aoki2019,Thomas2020,Thomas2021,Cairns2020}.
Here we demonstrate that UTe$_{2}$ single crystals displaying an optimal
superconducting transition temperature at $2$~K exhibit a single transition and
 remarkably high quality supported by their small residual heat capacity in the superconducting state and large residual resistance ratio. 
Our results shed light on the intrinsic superconducting properties of UTe$_{2}$ and bring into question whether UTe$_{2}$ 
is a multicomponent superconductor at ambient pressure.}

Uranium is a fascinating element located at the border between localized and delocalized $5f$ wavefunctions.
Uranium-based materials may therefore be found close to a magnetic-nonmagnetic boundary at which unconventional superconductivity is generally expected to emerge.
According to the Hill limit, superconductivity is favored when the distance between uranium atoms, $d_{U-U}$, is smaller than $3.6$~\AA, whereas localized wavefunctions favor magnetic order when $d_{U-U} > 3.6$~\AA~\cite{Hill1970, Moore2009}.
Unconventional actinide superconductors, however, remain a rather sparse class of strongly correlated materials 
that host many puzzling emergent properties.
Hidden order in tetragonal URu$_{2}$Si$_{2}$~\cite{Palstra1985, Mydosh2011}, time-reversal symmetry breaking in the superconducting state of hexagonal UPt$_{3}$~\cite{Schemm2014, Avers2020}, and contradicting reports on whether cubic UBe$_{13}$ is a spin-singlet or a spin-triplet superconductor~\cite{Han1986, Shimizu2015} are just a few examples. 
Other prominent examples include hexagonal antiferromagnetic U$M_{2}$Al$_{3}$ 
($M=$ Ni, Pd)~\cite{Geibel1991, Pfleiderer2009a} and orthorhombic ferromagnetic superconductors
UGe$_{2}$, UCoGe, and URhGe~\cite{Saxena2000, Aoki2019a}.

In 2019, orthorhombic UTe$_{2}$ became a new member of this family.
Early reports observed a superconducting transition at $T_{c} = 1.6$~K and a remarkably large upper critical field exceeding 40~T, a value much higher than the expected Pauli limit for a spin-singlet state~\cite{Ran2019, Ran2019a}.
Nuclear magnetic resonance (NMR) measurements found that the decrease in Knight shift below $T_{c}$ is much smaller than the expectation from spin-singlet pairing~\cite{Nakamine2019a}.
Though no magnetic order is observed above 25~mK $via$ muon spin resonance measurements~\cite{Sundar2019}, $a$-axis magnetization data can be described by the Belitz-Kirkpatrick-Votja theory for metallic ferromagnetic quantum criticality~\cite{Ran2019}.
UTe$_{2}$ was therefore proposed to be close to a ferromagnetic quantum critical point akin to UGe$_{2}$, UCoGe, and URhGe~\cite{Aoki2019a}.
The shortest U-U distance in UTe$_{2}$ within the $c$-axis dimers, $3.8$~\AA, supports proximity to a magnetic instability, but inelastic neutron scattering measurements as well 
as pressure- and field-dependent thermodynamic properties point to dominant antiferromagnetic fluctuations~\cite{Braithwaite, Thomas2020, Li2021, Duan2020, Duan2021, Knafo2021}.

Recently, the presence of two phase transitions in specific heat data combined with time-reversal symmetry 
breaking probed by the polar Kerr effect support the presence of a multicomponent superconducting order parameter in UTe$_{2}$.
Kerr trainability along the $c$ axis and symmetry requirements in the $D_{2h}$ space group further indicate that the two superconducting order 
parameters belong to a combination of either $B_{3u}$ and $B_{2u}$ or $B_{1u}$ and $A_{u}$ spin-triplet channels~\cite{Hayes2020}.
In this case, UTe$_{2}$ is a topological superconductor with Weyl nodes and surface Fermi arc states~\cite{Hayes2020}.

The observation of two superconducting transitions in UTe$_{2}$, however, remains disputed as independent groups observe a single transition~\cite{Cairns2020, Aoki2019, Thomas2021}.
The superconducting properties of UTe$_{2}$ are strongly dependent on the synthesis route, 
which further highlights the crucial role of sample quality in determining the intrinsic properties of unconventional superconductors. 
UTe$_{2}$ crystals grown by the self-flux method show no signs of bulk superconductivity, whereas crystals grown 
by chemical vapor transport show either a split transition or a single transition~\cite{Aoki2019}.
Notably, specific heat data show an apparent lack of entropy conservation between the superconducting and normal states, and a 
large residual Sommerfeld coefficient of unknown origin is observed in the superconducting state, $\gamma_{SC}$.
Further, the highest reported $T_{c}$ of 1.77~K yields a single transition and an inverse correlation between $T_{c}$ and $\gamma_{SC}$~\cite{Cairns2020,Aoki2020}.
Key outstanding questions are therefore whether the optimal $T_{c}$ in UTe$_{2}$ leads to entropy conservation and how the purported multicomponent transition responds to changes in $T_{c}$.

Here we show that UTe$_{2}$ crystals with the highest superconducting transition temperature, $T_{c} = 2$~K, exhibit a single thermodynamic 
transition. The high quality of the crystals is demonstrated by their high residual 
resistance ratio, RRR=57, and low residual heat capacity, $\gamma_{SC}$=23~mJ/mol.K$^{2}$, which leads to entropy conservation. Remarkably, 
normal state properties such as magnetic susceptibility and Sommerfeld coefficient remain
 unchanged between samples.
Lattice parameters and site occupancy determined from single crystal x-ray diffraction also do not change within experimental uncertainty.
Our results suggest that the superconducting state of UTe$_{2}$ is sensitive to remarkably subtle structural differences that 
deserve a central place in future investigations
of the intrinsic superconducting properties of UTe$_{2}$.

As described in Methods and summarized in Table I, single crystals of UTe$_{2}$ were grown using the chemical vapor transport (CVT) 
method with iodine as the transport agent.
Figure~1a shows the specific heat divided by temperature, $C/T$, as a function of temperature for seven representative samples. 
Sample s1 exhibits two well-defined features at 
$T_{c1} = 1.64$~K and $T_{c2} =1.48$~K,
which is consistent with results from Ref.~\cite{Hayes2020} (group 1) interpreted as distinct superconducting transitions from a 
multicomponent order parameter. Here $T_{c}$ is defined as the midpoint of the rise in $C/T$ on cooling.
This double feature is quickly replaced by a single
transition as the growth temperature decreases. Results for samples s2 ($T_{c} = 1.68$~K) and s3
($T_{c} = 1.77$~K) are consistent with reports from group 2 \cite{Aoki2019} and 
group 3 \cite{Cairns2020}, respectively. 

The optimal superconducting transition temperature is found in sample s6, whose bulk $T_{c}$
is $2$~K. The bulk superconducting transition quickly vanishes in crystals grown at even lower temperatures (sample s7). 
 Importantly, the residual heat capacity value in the superconducting state decreases monotonically as $T_{c}$ increases.
 Although changes in the U/Te 
 starting ratio were previously shown to affect $T_{c}$ \cite{Cairns2020}, our results demonstrate that the 
 optimal $T_{c}$ in UTe$_{2}$ is obtained at 
  lower growth temperatures. We find that slightly larger Te concentrations also quickly suppress $T_{c}$.
  In contrast to variations in $T_{c}$ and $\gamma_{SC}$, the normal state Sommerfeld coefficient is virtually constant for all samples,  
  $\gamma_{N} = 121(4)$~mJ/mol.K$^{2}$.

\begin{table}[h]
\caption{Crystal growth parameters and physical properties of UTe$_{2}$ single crystals.
$T_{i}$ ($T_{f}$) is the temperature of the hot (cold) end of the CVT temperature gradient.
}\label{Table1}
\begin{center}
\begin{tabular}{l@{\hspace{0.2cm}}c@{\hspace{0.2cm}}c@{\hspace{0.2cm}}c@{\hspace{0.5cm}}c@{\hspace{0.2cm}}c}								
		\hline
		\hline 													
     Sample          & $T_{i}$    & $T_{f}$  &	  $T_{c}$	          &	      $\gamma_{SC}$   & RRR   \\
                     & ($^{\circ}$C)     & ($^{\circ}$C)  &	  (K)	          &	       (mJ/mol.K$^{2}$)  &    \\
		\hline													
         s1	       &	1060	    &	1000	&	1.64 \& 1.48	          &	65	   & 30-40         	\\
         s2	       &	950	    &	860	&	1.68	          &	51	   &      -  	\\
         s3	       &	925	    &	835	&	1.77	          &	43	   &      -   	\\
         s4	       &	875	    &	785	&	1.85	          &	41	   & 50          	\\
         s5	       &	825	    &	735	&	1.95             &   25 &  	57	  	\\
         s6	       &	800	    &	710	&	2.00   		& 23	&	-\\		
         s7        & 775      & 685 & None       & - & 2 \\
	  \hline      					
	    \hline
\end{tabular}
  \end{center}
\end{table}

Figure~1b shows $C/T$ as a function of temperature for sample s6.
At $T_{c}$, the magnitude of the superconducting jump divided by the normal state Sommerfeld coefficient is $\Delta C/\gamma_{N}T_{c} = 1.8$. 
This value is larger than the weak coupling BCS limit of 1.43 and agrees with previous results on samples with a single transition higher than 1.7~K \cite{Cairns2020}. 
For samples with lower $T_{c}$, $\Delta C/\gamma_{N}T_{c}$ is smaller and ranges from $1.2$ to $1.5$ \cite{Ran2019,Hayes2020,Aoki2019,Thomas2020,Thomas2021}. 
Notably, a transition temperature of $\sim 2$~K has been observed previously in electrical resistivity data, but the associated bulk transition in $C/T$ 
occurred at lower temperature $\sim 1.77$~K~\cite{Cairns2020}. Whether the higher resistive transition is due to surface 
 effects or percolation through filaments in the bulk is still an open question.

\begin{figure}[b]
  \begin{center}
  \includegraphics[width=1\columnwidth]{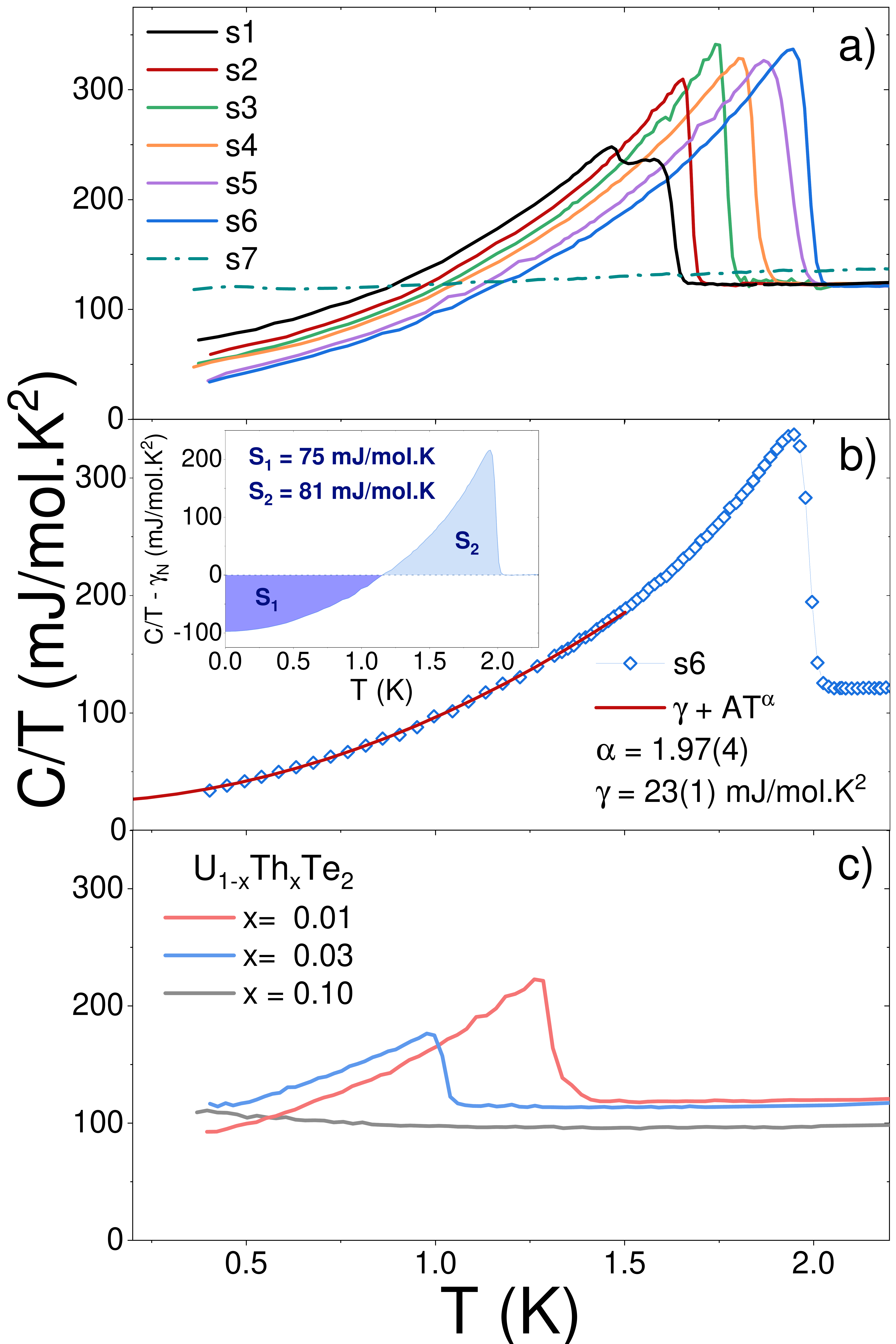}
  \vspace{-0.5cm}
  \end{center}
  \caption{Specific heat of UTe$_{2}$ single crystals. a) $C/T$ as a function of temperature for six representative samples. 
  b) $C/T$ as a function of temperature for sample s6. The solid line is a power-law fit below 1.5~K. Inset shows the entropy balance in a $C/T - \gamma_{N}$ $vs$ $T$ plot.
  c) $C/T$ as a function of temperature for Th-doped UTe$_{2}$ single crystals.}
  \label{fig:Fig1}
\end{figure}

The low-temperature $C/T$ behavior of sample s6 can be well described by the power-law expression $\gamma_{SC} + AT^{\alpha}$ wherein $\gamma_{SC} = 23$~mJ/mol.K$^{2}$ and $\alpha = 1.97 (4)$ (solid line in Fig.~1b).
The magnitude of the residual Sommerfeld coefficient in the superconducting state of sample s6 is the lowest reported value, which suggests that 
a larger $\gamma_{SC}$ value is not an intrinsic property of UTe$_{2}$.
In addition, the quadratic dependence of $C/T$ indicates the presence of point nodes, in agreement with previous thermal conductivity and 
specific heat measurements in crystals grown at higher temperatures~\cite{Metz2019,Kittaka2020,Cairns2020}.

The second order nature of the superconducting transition in UTe$_{2}$ requires entropy to be conserved at $T_{c}$. 
This equality can be probed by comparing the areas enclosed above and below the $\gamma_{N}$ baseline. The inset of Fig.~1b shows the difference between 
$C/T$ and $\gamma_{N}$ as a function of temperature as well as the corresponding areas $S_{1}$
and $S_{2}$. The magnitudes of the two areas differ by less than 8\%, in agreement with the expected entropy conservation 
in UTe$_{2}$, whereas samples with lower $T_{c}$ 
show an apparent entropy imbalance of about $60$\% \cite{Ran2019,Hayes2020,Thomas2020,Aoki2019}. The remaining small apparent entropy imbalance may be a hint 
that $T_{c}$ can still be further improved, though likely not by a significant amount. 
Alternatively, the imbalance could be tentatively explained by the presence of a nuclear Schottky anomaly at
lower temperatures. Finally, we note that a proper phonon subtraction 
was hindered by the fact that nonmagnetic analog ThTe$_{2}$ is not known to crystallize in the same strucutre of UTe$_{2}$. 

To test the solubility of Th in UTe$_{2}$, we investigate Th-doped UTe$_{2}$ single crystals grown in conditions 
similar to sample s1, which could also provide access to the unexplored regime of 
negative chemical pressure in UTe$_{2}$.
Figure~1c shows the specific heat divided by temperature as a function of temperature for U$_{1-x}$Th$_{x}$Te$_{2}$ at three Th concentrations.
At only 1\% Th doping, the superconducting anomaly is substantially suppressed by about 20\%. 
At such low doping, microprobe analysis using
energy dispersive x-ray spectroscopy shows that 
the actual concentration of Th is very close to the nominal concentration, and
 the doping is fairly homogeneous.
For the crystal shown in Fig.~1c, the mean actual concentration is 1.2\% and the 
homogeneity range is about 0.2\%.
At 3\% nominal Th doping, the mean actual concentration 
is $\sim$4\%, but a larger standard deviation of 2\% is observed within a crystal. 
The superconducting transition in specific heat is further suppressed
to $T_{c}=1$~K at 3\% nominal Th doping, whereas no transition is observed
at 10\% nominal Th doping. Microprobe analysis of the $x=0.1$ crystal shown in Fig.~1c
yields an actual concentration of 24(8)\%, but measurements in different crystals from the same batch
show significantly different dopings. These results are consistent
with an insolubitity region at larger Th concentrations.

Now we turn to the electrical and magnetic properties of sample s5, whose $T_{c}$ is 1.95~K. Figure~2 shows the anisotropic magnetic 
susceptibility, $\chi(T)$, of UTe$_{2}$
 as a function of temperature. 
 Importantly, $\chi(T)$ in the normal state is identical to previous reports \cite{Ran2019,Aoki2019} and 
between different samples in this work. The a-axis susceptibility is the largest, which
suggests that the $a$ axis is the easy axis. 
The $c$-axis susceptibility is small and monotonic, whereas the 
$b$-axis susceptibility shows a broad feature centered at $\sim 35$~K. 
The right inset of Fig.~2 displays the zero-field-cooled and field-cooled 
$\chi(T)$ at 2~Oe with field applied along the $a$ axis. A clear
diamagnetic signal sets in at $1.95$~K, which is consistent with
electrical resistivity and specific heat data.

\begin{figure}[!ht]
  \begin{center}
  \includegraphics[width=1\columnwidth,keepaspectratio]{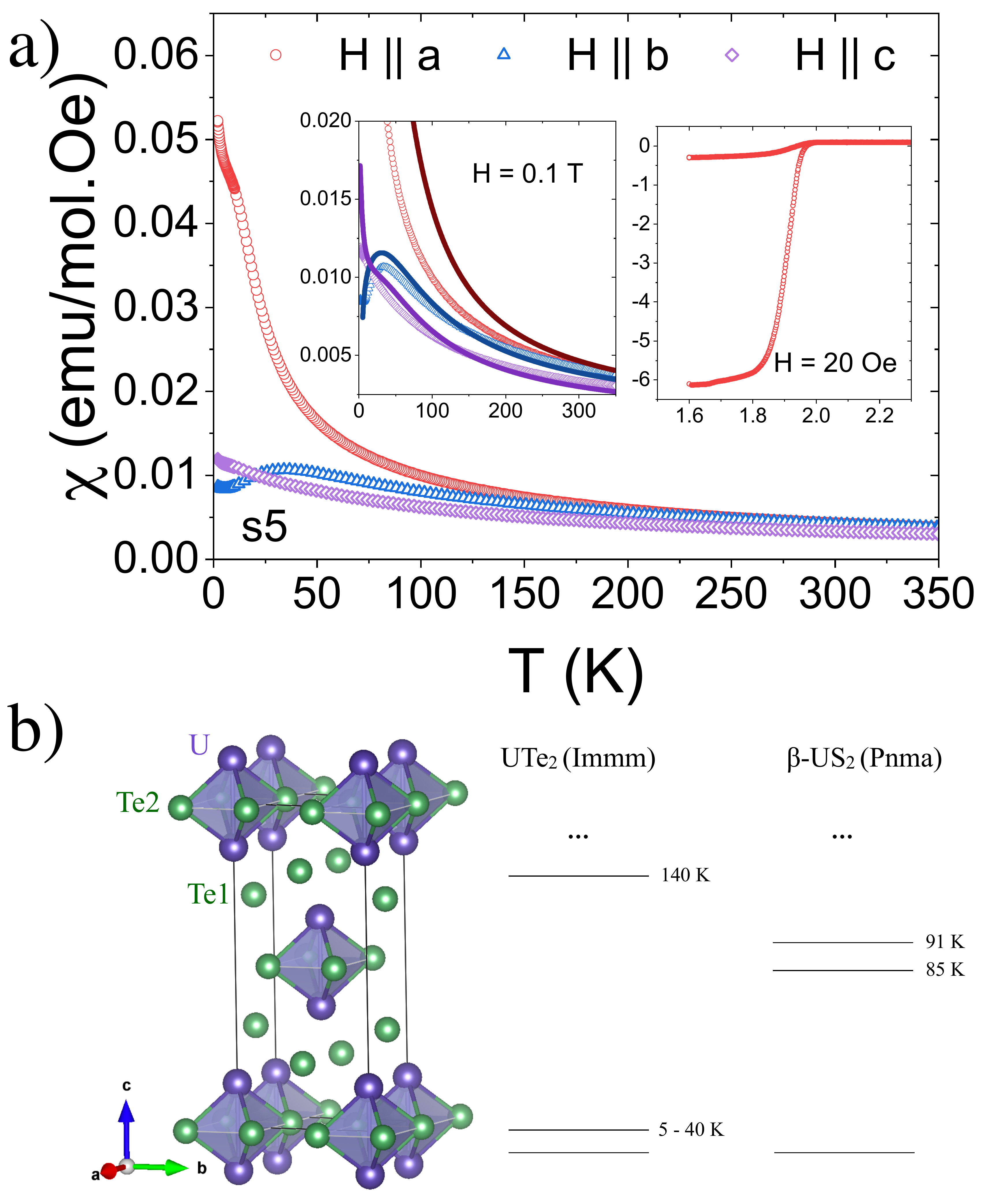}
  \end{center}
  \vspace{-0.3cm}
  \caption{a) Anisotropic magnetic susceptibility of UTe$_{2}$ as a function of temperature at 0.1~T. Right inset shows the zero-field-cooled and field-cooled 
  magnetic susceptibility at 2~Oe with field applied along the $a$ axis. Left inset shows the high-temperature anisotropic magnetic susceptibility
  and the associated crystal electric field fits (solid lines). b) (Left) Crystal structure of UTe$_{2}$ highlighting the polyhedra enclosing the 
  $c$-axis uranium-uranium dimer. (Right) Comparison of crystal electric field levels of UTe$_{2}$ and $\beta$-US$_{2}$.}
  \label{fig:Fig2}
\end{figure}

Crystal electric field (CEF) effects are able
to capture the qualitative $\chi(T)$ behavior of UTe$_{2}$. The solid lines in the left
 inset of Figure~2
show fits of the data to an orthorhombic CEF Hamiltonian
$\mathcal{H}_{CEF}=B_{2}^{0}O_{2}^{0} + B_{2}^{2}O_{2}^{2} + B_{4}^{0}O_{4}^{0} + B_{4}^{2}O_{4}^{2} + B_{4}^{4}O_{4}^{4}$,
where $B_{i}^{n}$ are the CEF parameters, and $O_{i}^{n}$ are the Stevens equivalent operators obtained from the angular momentum operators~\cite{Pagliuso2006}.
Here we consider the $5f^{2}$ configuration of uranium, $i.e.$, U$^{4+}$ ($J=4$, $S=1$), as the localized configuration that gives rise to CEF effects.
This consideration is based on two experimental results. First, x-ray absorption measurements under pressure show that UTe$_{2}$ is mixed valence at ambient pressure
and goes towards $4+$ when magnetic order sets in under pressure \cite{Thomas2020}. Second, core-level spectroscopy supports the mixed valence nature of UTe$_{2}$ 
wherein the dominant contribution arises from the 
itinerant $5f^{3}$ configuration and a smaller localized $5f^{2}$ contribution is responsible for a satellite peak \cite{Fujimori2021}.

The orthorhombic crystalline environment splits the 9-fold degenerate multiplet of $J=4$ U$^{4+}$ into a collection of singlets. The relevant levels 
below room temperature are described by a combination of two low-lying singlets and 
an excited singlet at 140~K. As shown in Figure~2b, this configuration resembles that of $\beta$-US$_{2}$, whose experimentally-determined 
crystal field levels are given by a ground state singlet separated by 85~K and 91~K from two excited singlets. 
Akin to UTe$_{2}$, $\beta$-US$_{2}$ also orders magnetically under pressure, which indicates that the 
admixture of three low-lying singlets yields a finite magnetic moment. In fact, the ground state and the second 
excited state at 140~K in UTe$_{2}$ form a quasi-doublet, $i.e.$, they share the same $|j_{z}\rangle$ contributions $|\pm 4\rangle$, $|\pm 2\rangle$, and $|0\rangle$.
The CEF parameters and corresponding energy levels and wavefunctions are shown in Table~S1 (Supplemental Information).

Figure~3 shows the electrical resistivity with applied current along the $a$ axis, $\rho_{100}$, as a function of temperature. 
At high temperatures, $\rho_{100}$ increases slightly on cooling, which is consistent with 
previous reports and stems from incoherent Kondo scattering.
At about $40$~K, $\rho_{100}$ decreases sharply on cooling, a behavior typically attributed to the formation of a Kondo coherent state.
This coherence temperature is also consistent with estimates from scanning tunneling microscopy~\cite{Jiao2019}.

\begin{figure}[!h]
\begin{center}
\includegraphics[width=1\columnwidth]{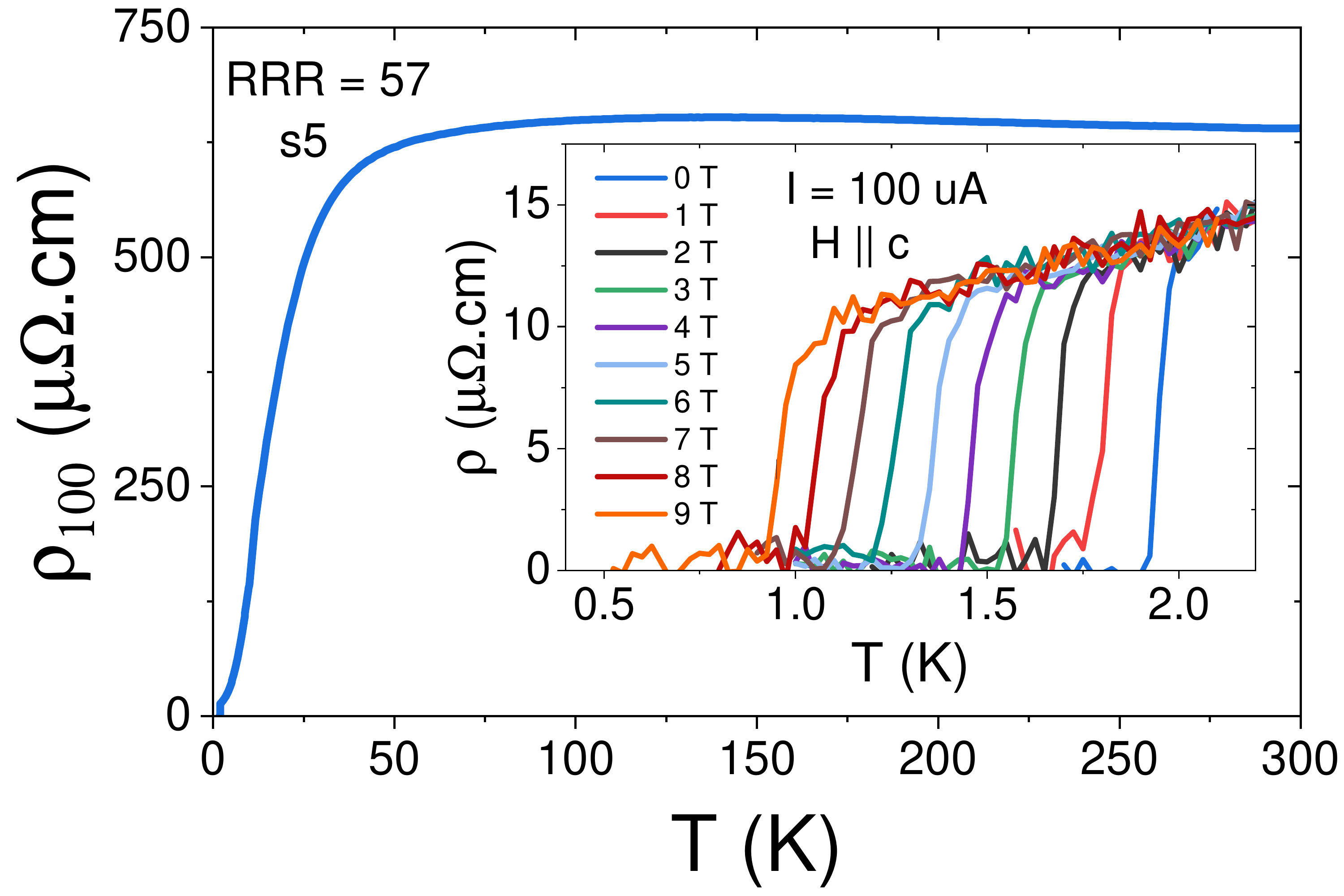}
\end{center}
\vspace{-0.5cm}
\caption{Electrical resistivity of UTe$_{2}$ as a function of temperature with current along the $a$ axis. 
Inset shows the low-temperature behavior under magnetic fields applied along $c$.} 
\vspace{-0.2cm}
\label{fig:Fig3}
\end{figure}

The inset of Fig.~3 shows the low-temperature behavior of $\rho_{100}$
at various magnetic fields applied along the $c$ axis.
At zero field, the mid-point of the superconducting transition is at 1.95~K, 
which is precisely the value obtained from specific heat measurements.
At 9~T, $T_{c}$ is reduced to 1~K, which is consistent with previous reports
by taking into account a $0.3$~K shift in the zero-field $T_{c}$~\cite{Ran2019, Aoki2019}.
Finally, the residual resistivity ratio (RRR), defined as $[\rho(300\:K) - \rho(T = 0))/\rho(T = 0)]$, 
is 57, which is the highest reported value for $\rho_{100}$.
In contrast, the RRR value of non-superconducting sample s7 is only 2 (see Table 1).
The residual resistivity, $\rho(T = 0)= \rho_{0} = 11 \mu\Omega$.cm, was obtained by a low-temperature fit
to $\rho_{0} + AT^{2}$, and to our knowledge is also the lowest reported value.

Both $\rho_{0}$ and RRR values 
are commonly used criteria for the presence of disorder and have been successfully utilized to infer
the quality of unconventional superconductors, including UTe$_{2}$ by groups 1 and 2 \cite{Hayes2020,Aoki2019}.
The pressing question therefore relates to the cause of the underlying disorder in UTe$_{2}$. Historically, planar defects, grain boundaries, and substitutional or interstitial impurities have been argued to affect the sample quality of various actinide superconductors including UPt$_{3}$~\cite{Kycia1998}, UBe$_{13}$ \cite{UBe13}, UCoGe~\cite{Huy2009}, and URu$_{2}$Si$_{2}$~\cite{Gallagher2016}. 
More broadly, disorder has been shown to reduce $T_{c}$ in other unconventional superconductors 
such as Sr$_{2}$RuO$_{4}$ \cite{SRO} and FeSe \cite{FeSe}. 
Recent reports have argued that Te vacancies are responsible for lower superconducting
transitions in UTe$_{2}$ \cite{Cairns2020}. Remarkably, in the present study we do not observe statistically relevant differences in microprobe analysis through energy dispersive x-ray spectroscopy.
All single crystals investigated here showed a stoichiometry of UTe$_{2.2(3)}$, $i.e.$, the large error bars hinder 
the establishment of any possible trends.
This result is supported by standard laboratory single crystal x-ray diffraction of samples s1 and s6, wherein
both uranium and tellurium sites are fully occupied. 
In addition, lattice parameters as well as all refined parameters are constant across all investigated samples within experimental uncertainty. 
Table S2 (Supplemental Information) provides details of the full refinements.
Our results suggest
that the superconducting state of UTe$_{2}$ is remarkably sensitive to disorder and calls attention to the 
importance of determining the main structural parameter that suppresses and splits $T_{c}$.


In summary, we report the optimal superconducting transition temperature, $T_{c} = 2~$K, in UTe$_{2}$ single crystals. 
Our crystals exhibit a single superconducting transition 
and their high quality is demonstrated by high residual 
resistance ratios, RRR~$=57$, and low residual heat capacity values in the superconducting state, $\gamma_{SC} = 23$~mJ/mol.K$^{2}$, which leads the expected entropy conservation.
The correlation between $T_{c}$ and residual resistance ratios underscores the role of disorder 
in the superconducting state of UTe$_{2}$.
The disappearance of the double transition feature as sample quality is improved brings into question whether
 UTe$_{2}$ is a multi-component superconductor at ambient pressure.

We would like to acknowledge constructive discussions with M Bordelon, N Butch, JP Paglione, A Huxley, and RM Fernandes.
This material is based upon work supported by the U.S. Department of Energy, Office of Science, National Quantum Information Science Research Centers, Quantum Science Center. 
Scanning electron microscope and energy dispersive x-ray measurements
were supported by the Center for Integrated Nanotechnologies,
an Office of Science User Facility operated for
the U.S. Department of Energy Office of Science.
AW acknowledges support from the Laboratory
Directed Research and Development program at LANL.

\subsection{Methods}

Single crystals of UTe$_{2}$ were grown using the chemical vapor transport method.
Solid pieces of depleted uranium ($99.99$\%) and tellurium (Alfa Aesar, $99.9999+$\%) were weighed in a 2:3 ratio
with total mass of $\sim 1$~g. The elements were sealed under vacuum using a hydrogen torch in a quartz tube along with
 $\sim 0.2$~g of iodine (Alfa Aesar, $99.99+$\%).
The dimensions of the quartz tube are 1.8~cm (outer diameter), 1.4~cm (inner diameter), and
$\sim 15$~cm (length), which resulted in an iodine density of about $0.8$~mg/cm$^{3}$.
A temperature gradient was maintained in a multi-zone furnace for 11 days. 
The elements were placed in the hot end of the gradient at $T_{i}$, whereas single crystals of 
UTe$_{2}$ were obtained at $T_{f}$, the cold end of the gradient. $T_{i}$ was varied from 1060~$^{\circ}$C 
to 800~$^{\circ}$C, whereas $T_{f}$ was varied from 1000~$^{\circ}$C 
to 710~$^{\circ}$C. A summary of the growth conditions of representative samples is presented in Table~1.
For Th-doped samples, Th and U were arc melted in a water-cooled Cu hearth prior to the growth.

The crystallographic structure of UTe$_{2}$ was determined at room temperature by a Bruker D8 Venture single-crystal diffractometer equipped with Mo radiation. 
Elemental analysis of our single crystals using energy-dispersive x-ray spectroscopy in a commercial scanning electron microscope. 
Single crystals of UTe$_{2}$ are sensitive to air and moisture, and they were kept in an argon glovebox between measurements to allow for sample stability 
over several months.  Magnetization measurements were obtained through a commercial SQUID-based magnetometer. Specific heat measurements were made 
using a commercial calorimeter that utilizes a quasi-adiabatic thermal relaxation technique. The electrical resistivity ($\rho$) was 
characterized using a standard four-probe configuration with an AC resistance bridge.
Values of RRR in Table I were determined for current flow along the $a$ axis.

\end{document}